\documentclass[preprint,preprintnumbers,amsmath,amssymb,longbibliography]{revtex4-1}
\makeatother
\usepackage[dvips]{graphicx}
\usepackage{lmodern}

\usepackage{amsmath}
\usepackage{amsbsy}
\usepackage{caption}
\usepackage{color}
\DeclareGraphicsExtensions{.pdf,.eps,.png,.jpg,.mps}
\usepackage{gensymb}

\definecolor{textcolor}{cmyk}{0,0,0,1}
\definecolor{magenta}{rgb}{1,0,1}
\definecolor{green}{rgb}{0,1,0}
\definecolor{red}{rgb}{1,0,0}

\begin{document}

\title{
An Array of Layers in Silicon Sulfides:
Chain-like and Ground State Structures
}
\author{ T. Alonso-Lanza,$^{1}$ F. Aguilera-Granja,$^{1,2}$ A. Ayuela$^{1}$}
\affiliation{
$^1$Centro de F\'{\i}sica de Materiales CFM-MPC CSIC-UPV/EHU, Donostia
International Physics Center (DIPC), Departamento de F\'{\i}sica de Materiales, Fac. de Qu\'{\i}micas, UPV-EHU, 20018 San Sebasti\'an, Spain
\\
$^2$Instituto de F\'{\i}sica, Universidad Aut\'onoma de San Luis de Potos\'{\i}, 78000 San Luis Potos\'{\i} S.L.P., M\'exico
\\
}


\begin{abstract}
\textbf{While much is known about isoelectronic materials related to carbon nanostructures, such as boron nitride layers and nanotubes, rather less is known about equivalent silicon based materials. Following the recent discovery of phosphorene, we herein discuss isoelectronic silicon monosulfide monolayers.  We describe a set of anisotropic ground state structures that clearly have a high stability with respect to the near isotropic silicon monosulfide monolayers.
 The source of the layer anisotropy is related to the presence of Si-S double chains linked by some Si-Si covalent bonds, which lye at the core of the increased stability, together with a remarkable {\it spd} hybridization on Si.
The involvement of {\it d} orbitals brings more variety to silicon-sulfide based nanostructures that are isoelectronic to phosphorene, which could be relevant for future applications, adding extra degrees of freedom.}

\end{abstract}

\maketitle
The isolation of a single layer of carbon atoms in 2004\cite{novoselov2004electric} opened up a new and exciting field in science related to the study and characterization of monolayers, which is still a growing field today. The impresive electronic, optical and mechanical properties of graphene\cite{novoselov2012roadmap,huang2011graphene,geim2009graphene,neto2009electronic,bolotin2008ultrahigh,zhang2005experimental} were a key motivation in the search for new monolayers, which has the aim of discovering nanomaterials that do not have the drawbacks that graphene has in some applications, such as the absence of a band gap; alternatively some new and unexpected properties caused by bidimensionality may be possible.
The growing interest in graphene is currently driving the development of experimental techniques needed to characterize single layer materials with sufficient accuracy\cite{hashimoto2004direct,liu2009open,jin2009deriving}. This improvement, together with the possibility of exfoliating weakly bound layered materials, explains the proliferation of interest in new two-dimensional layers.
Among the isoelectronic compounds of graphene are boron nitride and silicene. Boron nitride nanotubes were synthesized\citep{chopra1995boron} within the carbon row of the periodic table, and hexagonal boron nitride layers were later exfoliated from bulk in the most stable phase which is built from weakly bound monolayers \cite{zunger1976optical}.
Lower down in the same group of the periodic table group, silicene has been synthesized experimentally some years ago\cite{vogt2012silicene,lalmi2010epitaxial,aufray2010graphene}.
Silicene has some important differences from graphene, including a buckled structure, higher reactivity and a more easily tunable gap by surface adsorption\cite{ni2014tunable,du2014tuning}. All these silicon-type properties stem for the {\it sp$^3$} hybridization and the double bond rule\cite{gusel1979formation,jutzi1975new,mulliken1950overlap}, which state that double bonds are not formed for elements in the third period such as silicon. Other Group IV structures similar to graphene and graphane have been proposed\cite{garcia2011group}.
Nevertheless, the field broadened to systems noticeably different from graphene. Our interest lie in the search for isoelectronic compounds related to other recent synthesized monolayers, but which still have silicon with its special properties as a principal ingredient.

A new impetus has recently come from the study of single layers of black phosphorus\cite{liu2014phosphorene,castellanos2014isolation}, termed phosphorene in analogy with graphene and having desirable characteristics for electronics, including high carrier mobility, anisotropic electronic properties, and a band gap that depends on thickness\cite{xia2014rediscovering}. Despite some drawbacks in terms of fast degradation on contact with air\cite{wood2014effective,island2015environmental}, phosphorene shows promise for applications such as field-effect transistors\cite{li2014black,castellanos2015black}.
For this reason, the investigations of  isoelectronic compounds of phosphorene is currently of great interest.
The same line of enquiry has lead to the proposal of isoelectronic monolayers composed of Group V elements made either of single elements, such as arsenic and antimony monolayers known as arsenene and antimonene\cite{kamal2015arsenene,zhang2015atomically}, or mixed together as AsP and SbP \cite{weiyang}.
Furthermore, the significance of boron nitride (with respect to graphene) shows that is logical to consider compounds of Groups IV-VI, also called Group IV monochalcogenides\cite{gomes2015enhanced,tuttle2015large,mehboudi2015two}, which are also isoelectronic to phosphorene as well as being semiconductors with band gaps larger than those in the bulk phase\cite{singh2014computational}. Among these, our attention was drawn particularly to silicon monosulfide. Monolayers of this material have been reported to display two structures close together in energy, the more stable of these being similar to graphene and the other less stable being similar to phosphorene\citep{zhu2015designing}. Although both forms contain silicon atoms bonded to three different sulfur atoms, silicon compounds can present higher coordination sometimes even being pentacoordinated
\citep[and references therein]{ayuela2007silicate}. We must therefore allow for the greater coordination of silicon atoms within silicon monosulfide layers when considering the range of possible forms.

In this work we present a new anisotropic $\gamma$ structure for the silicon monolayer that has greatly improves stability compared with previous reports of two dimensional layers. We begin by considering the geometry of the new ground state, which contains silicon-silicon bonds while each silicon atom also remains bonded to three sulfur atoms. In our simulations we found that this anisotropic and more stable $\gamma$ layer maintains its structural integrity at room temperature. We analyzed the bonding and the hybridization of the sulfur and silicon atoms and found that the silicon-silicon bond seems to behave like a double bond, in violation of the double bond rule. More importantly, the coordination shown by the silicon atoms is strongly influenced by $spd$ hybridization. This finding paves the way using large cells for the search of a whole array of two dimensional structures that are both stable and highly anisotropic, all of which could be of interest, for instance in the patterning of nanostructures with specific properties on various substrates.

\section*{Results and discussion}

\subsection*{Monolayers}

\begin{figure*}[thpb]
      \centering
\includegraphics[clip,width=\textwidth,angle=0,clip]{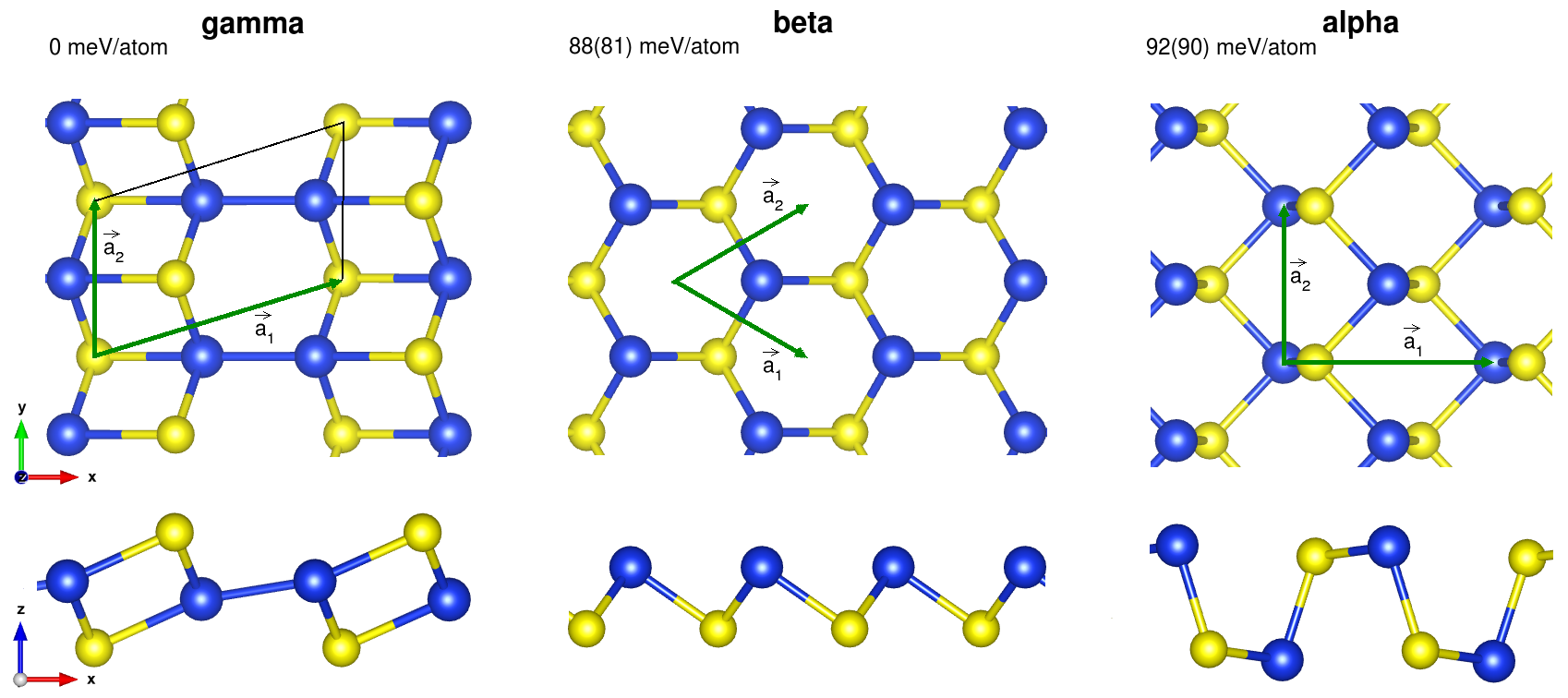}
\caption{\label{figure1}
\textbf{Geometries of three silicon monosulfide monolayers in order of increasing energy.} Energies in parenthesis correspond to test results obtained using the VASP code. The unit cell employed for the electronic analysis is shown in each case. Silicon (sulfur) atoms are denoted by blue (yellow) spheres; this color code is used throughout the article. The $\gamma$ layer is the most stable and anisotropic of the three forms.}
\end{figure*}

\subsubsection*{Anisotropy in highly stable structures}

Figure \ref{figure1} shows three different monolayer arrangements. We
obtained a monolayer labeled $\gamma$ which improves the binding energy per atom by 88 meV with respect to the recently proposed $\alpha$ and $\beta$ structures \cite{zhu2015designing}.
We note that this ground state is more stable than other forms according to two different codes: the VASP values are shown in parentheses in Fig. \ref{figure1}. The previously reported second and third most stable monolayers, $\alpha$ and $\beta$ were obtained using a restriction to three coordinated elements in either isotropic hexagonal or near-isotropic rectangular (close to square) structures. Our results are in fully agreement  with this previous study, although we found $\beta$ to be more stable than $\alpha$, by 4meV instead of the reported 12meV, because the distances for the $\alpha$ monolayer are slightly different.
More importantly, we found the $\gamma$ ground state for the SiS monolayer because the unit cell was assumed rhombohedral, and it is crucial that it contains two silicon and two sulfur atoms.  Silicon monosulfide monolayers therefore present a new ground state when larger cells are allowed.

Although the structure of black phosphorous is known to have some underlying anisotropy, this is clearly revealed in the $\gamma$ low lying stable layer when larger cells are allowed.   Figure \ref{figure1}  shows alternating parallel rows of rippled hexagons and squares along the y direction. Matching these two patterns, silicon atoms adopt a four-fold coordination, where they bind to three sulfur atoms in the squares and  more interestingly form Si-Si dimers ribbing the hexagons at the sides in a stair-like fashion.

We then applied molecular dynamics using the Nose thermostat  to study the stability of the $\gamma$ monolayers.
The data used for the dynamics are set out in Supplementary Information where we consider the energy and position according to the number of time steps. The Si and S atoms fluctuate around their relaxed equilibrium positions and we show that the $\gamma$ monolayer is stable up to room temperature.

\subsubsection*{spd hybridization in Si bonds}

We now look in detail at the bonding parameters with the aim of clarifying the mechanism that explains the stability order. Firstly, we note that $\beta$ and $\alpha$ monolayers each have a silicon bonded to three sulfur atoms, with three distances for the isotropic $\beta$, two of 2.37\AA{} and one for the almost isotropic $\alpha$ of 2.39\AA{}. The analysis is completed by measuring the angles between the established bonds, which are 90.$\degree$ for $\beta$ and one of 93.$\degree$ and two of 96.6$\degree$ for $\alpha$.
However, the silicon atoms in the $\gamma$ monolayer present four bonds instead of three, which constitute a significant difference with respect to the structures described above. Three bonds are still made with sulfur atoms of about 2.31\AA{}, and the extra one links two silicon atoms separated by 2.50\AA{}.  The angles between these bonds are relevant for this kind of implied hybridization: there are two of 92 $\degree$, and 94$\degree$ for the case of the silicon-sulfur bonds, and two of 99 $\degree$ and 164$\degree$ for the angles between the silicon-silicon bond and the other three silicon-sulfur bonds, respectively.
These angles reveal the symmetry adopted by the orbitals of the silicon atom, which is closely related to {\it spd} hybridization. The involvement of $d$ electrons on Si seems key to explaining the bond in the SiS layers.

\begin{figure*}[thpb]
      \centering
\includegraphics[clip,width=\textwidth,angle=0,clip]{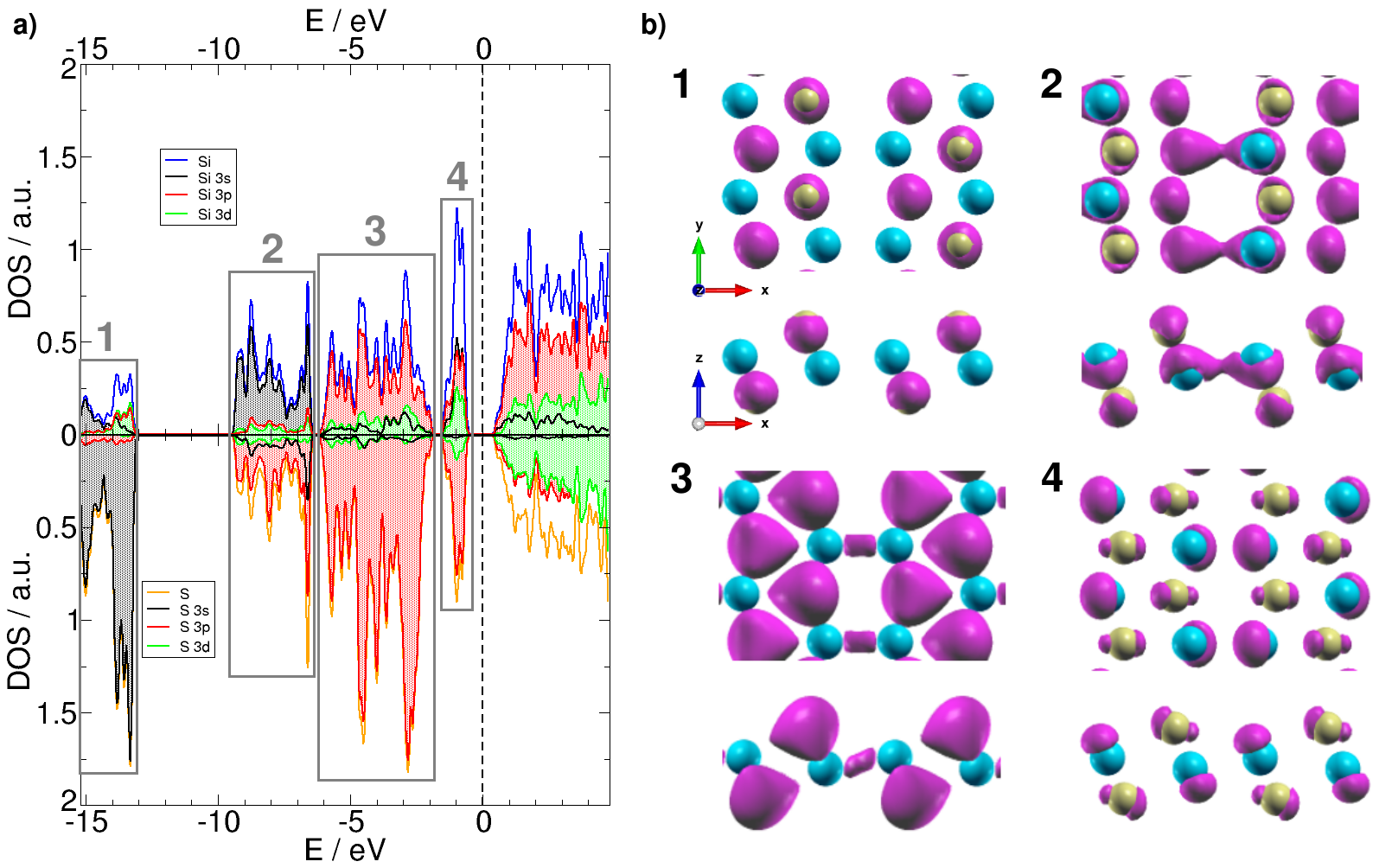}
\caption{\label{figure2}
\textbf{Electronic structure of $\gamma$ monolayer.} (a) Projected density of states. The four zones 1-4 highlighted within the rectangles are separated by small gaps. (b) Local density of states in energy zones 1-4 shown by lobules. Zones 2-3 have a component in the SiSi bond. Zone 4 is p on sulfur and a dangling bond on Si, with a large $spd$ hybridization.
}
\end{figure*}

In order to asses the hybridization in more detail we consider the electronic structure of the $\gamma$ layer. The density of states (DOS) projected for the silicon and sulfur orbitals are shown in Fig.\ref{figure2}(a). The total DOS below the Fermi energy is composed of four distinct zones separated by small gaps in energy. The rectangles in Fig.\ref{figure2} enclose the four different zones in terms of energy and we detail the orbital contribution further.
Notice that the contribution of sulfur is larger than that of silicon for almost all zones because of the high number of valence electrons. More relevant is the fact that in zone 4, the peak close to the Fermi energy has a preeminent contribution from silicon. This zone 4 is formed by {\it s} and {\it p} Si orbitals in equal proportion and from {\it d} at half proportion, which is evidence of the {\it spd} hybridization.

At deeper energies we find consecutively two zones, 3 and 2, that are mainly contributed by {\it p} and {\it s} Si orbitals respectively.
Looking at the sulfur orbitals, we find {\it s} states deep in energy in zone 1 but scarcely hybridized with {\it p} orbitals in other zones, which constitutes almost all the contribution of sulfur at higher energies,  now with a much smaller {\it d} part than at the highest occupied energy.
We illustrate the energy zones in space using the localized density of states in Fig.\ref{figure2}(b). We find that the silicon-silicon bond comes mainly from $\sigma$- and $\pi$-like states in zones 2 and 3, respectively,  lying lower than the Fermi energy. This can explain the enhanced stability of the $\gamma$ monolayer.
This seems to be evidence of a double bond between two silicon atoms, adding to the  body of few known cases where the double bond rule is not satisfied.
As previously commented, the role of this bond is to enhance stability in the SiS layers, which clarifies its importance.

It is noteworthy that just below the Fermi energy, in zone 4, the
spatial LDOS shape is {\it p}-like  over sulfur atoms and has lobules on silicon stemming from the hybridization of {\it spd} electrons, in agreement with the above results. These lobules on silicon represent something like a dangling bond. Even though both zones 3 and 4 have {\it p} contributions, we note that by looking at the side views of zones 3 and 4 that they are largely orthogonal to each other.

\subsubsection*{Anisotropy in band structures}

\begin{figure*}[thpb]
      \centering
\includegraphics[clip,width=\textwidth,angle=0,clip]{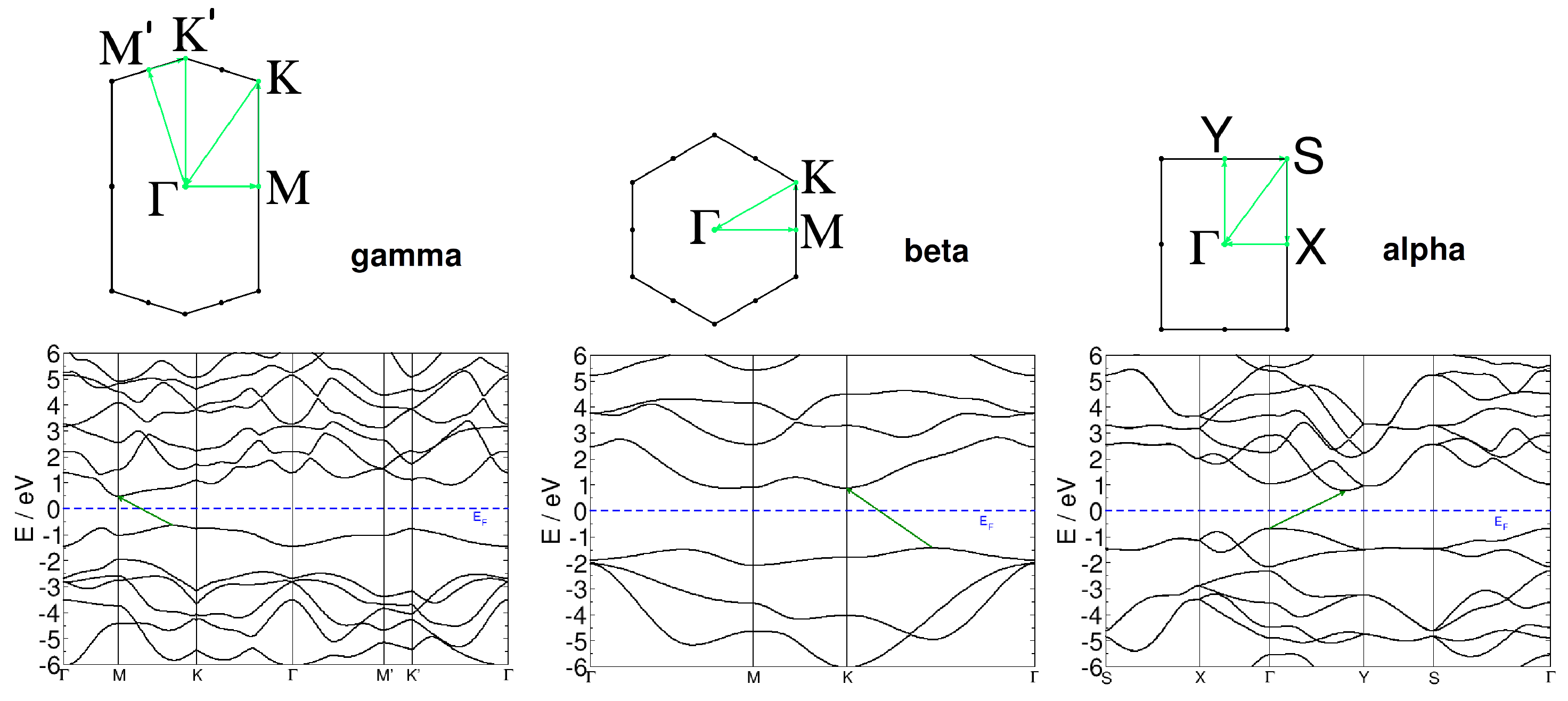}
\caption{\label{figure3}
\textbf{Band structures.} The band structures of three silicon monosulfide monolayers together with the path followed in the reciprocal space for each case.}
\end{figure*}

We next comment in detail on the implications of the anisotropy on the electronic band structures, as shown in Fig.\ref{figure3} for the three different monolayers. The band structures we obtained for $\alpha$ and $\beta$ layers faithfully reproduce those presented previously, and the band gaps are 1.5 eV and 2.3 eV respectively \cite{zhu2015designing}. A common feature among all of them is the presence of an indirect band gap.  The band gap for $\gamma$, the ground state monolayer, decreases to 1.13 eV, the smaller value of the three monolayers, so the assumed correlation between the gap and the stability does not apply.
\footnote{ 
Consideration of the charge transfer under the Mulliken scheme represents another step towards further understanding of the electronic structure of the new ground state monolayer. As expected due to the larger electronegativity of sulfur with respect to that of silicon 0.4 electrons pass from the silicon to the sulfur atom. For $\alpha$ and $\beta$ we find values of 0.46 and 0.33 electrons respectively, which are slightly high than those given by Zhu et al\cite{zhu2015designing}, of 0.3 and 0.2 electrons respectively. It appears that for the three monolayers both the trend and the amount of charge transfer are similar. Considering the new ground state $\gamma$, this does not present noticeable differences that can explain the gain in stability. Consequently, a new approach is required  to identify the reason behind the enhanced binding energy per atom such as the {\it spd} hybridization.}
 It should be noted that the presence of an isolated valence band close  to the Fermi energy  for the $\gamma$ layer is a feature that distinguishes it from the $\alpha$ and $\beta$ ones.
The origin of this band was clarified above as being due to $p_z$-like sulfur orbitals and {\it spd} Si lobes.
Furthermore, for the $\gamma$ monolayer the rhombohedral unit cell implies a distorted hexagonal Brillouin zone. The band is anisotropic in the plane and the valence-conductance distance is lower around the M-K point, where there is a strong y component along the Si dimers. They seem crucial not only in terms of stability, but also for the peculiar anisotropic properties in holes or electrons, which could be important for future applications of SiS monolayers in the design of possible conductance devices.

\subsection*{An array of structures}

\begin{figure}[thpb]
      \centering
\includegraphics[clip,width=0.5\textwidth,angle=0,clip]{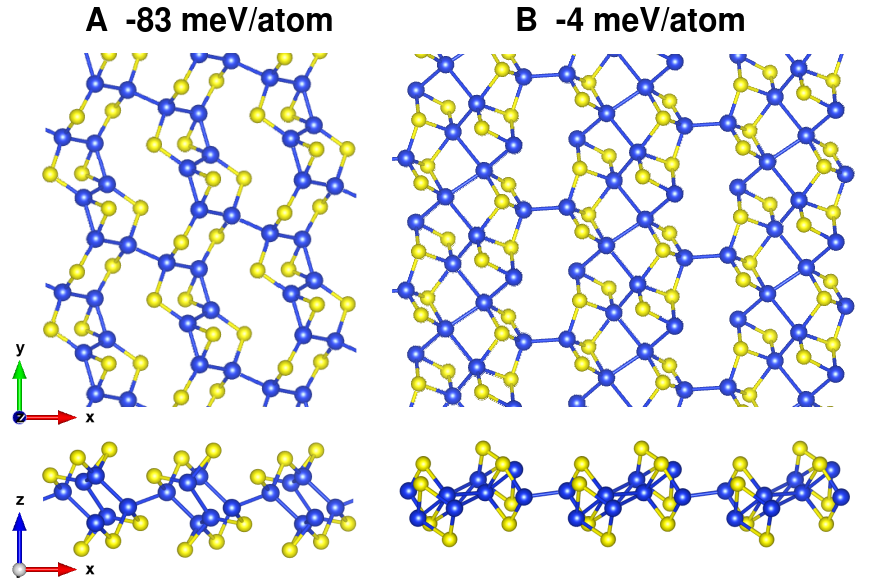}
\caption{\label{figure4}
\textbf{More SiS chain-like structures.} Examples of stable monosulfide chain-like structures linked by Si dimer, obtained with large cells. The reference for energies is referred to the $\gamma$ monolayer from Fig.\ref{figure1}. Note that Si atoms form larger chains, percolating through the SiS layer. A further two different structures are shown in the Supporting Information.}
\end{figure}

In addition to the findings outline above, we found other anisotropic structures with chain like geometries where the number of atoms per cell is much greater than in the  $\gamma$, $\alpha$ and $\beta$ structures described above. As input structures we  generated a great variety of bipartite lattices with different symmetries such as square, rectangular, and hexagonal, giving the possibility of steps of different heights. All these initial models were then optimized by relaxing positions and cell lattice vectors. In Fig.\ref{figure4} we display two of the most stable structures we have found thus far. These structures present a long-range spatial order and are more stable than the regular ones given above: one of them is even preferred to the $\gamma$ monolayer by 83 meV/atom.
In line with the findings for the $\gamma$ monolayer, the structures present the same pattern: more or less complicated chains of Si-S repeated and linked by silicon-silicon covalent bonds between adjacent chains. The distance between these Si-Si bonds is around 2.4\AA{}, denoting strong bonding as for the $\gamma$ monolayer.

We found other stable structures such as those shown in Fig.\ref{figure4} competing in energy with the $\gamma$ monolayer, established as the zero of energy. The cross sections of Fig.\ref{figure4} shows that those new structures are typically thicker.  The number of atoms per surface unit is increased, which could add to the enhancement of stability with respect to the monolayers, with lower densities with thicknesses of barely two atoms. 
Another difference is the increase in the number of silicon atoms bound together, with Si linear chains up to four atoms, which seems to percolate through SiS layers. The presence of silicon-silicon bonds within a chain lowers the coordination of the sulfur atoms in order to maintain the stoichiometry of SiS, which produces holes in the structure favoring the wire in the long-range structure.
The number of new structures suggests the richness of the different configurations that may occur due to different hybridizations in the electronic structures, clearly different to graphene, particularly in the case of those elements closely isoelectronic to the phosphorene group and in the same row of the periodic table.

\section*{Summary and conclusions}

In summary, we have presented a new ground state monolayer labeled as $\gamma$ for silicon monosulfide. It has enhanced stability with respect to the well-known almost isotropic states. Following the search for the $\gamma$ layer we also found several forming an array of silicon monosulfide thicker layers. All of these structures present chains with a long range spatial order and silicon-silicon bonds linking them,  either more stable than the $\gamma$ monolayer or competing with it.
 The silicon atoms participate in direct Si-Si double bond where {\it s}, {\it p} and {\it d} orbitals hybridize, like {\it spd}. The extra stability for the $\gamma$ monolayer and the related structures comes from the extra silicon-silicon bonds,  that were not present in the other silicon monosulfide monolayers, and brings forward the structure and electronic anisotropy of these layers.  It seems that the anisotropy in structures and bands isoelectronic to phosphorene differs from compounds isoelectronic to graphene, and can be important for hole and electron conductance in SiS monolayers.

\section*{Computational details}
We performed calculations on silicon sulfide nanostructures using the SIESTA (Spanish Initiative for Electronic Simulations with Thousands of Atoms) method making reference to density functional theory.   For the exchange and correlation potentials, the Perdew-Burke-Ernzenhof form of the generalized gradient approximation (GGA) \cite{perdew1996generalized} is well suited to the study of these kinds of covalent nanolayers in combination with the description of atomic cores by non-local norm-conserving Troullier-Martins\cite{troullier1991efficient} pseudopotentials factorized in the Kleynman-Bylander form.
Double zeta plus polarization orbitals basis set for valence electrons.
The same computational parameters were used in all the calculations, namely an electronic temperature of 25 meV and a meshcutoff of 250 Ry. We sample the Brillouin zone using a kgrid cuttoff of 25 \AA{}.
All cells were assigned large vectors (24.5\AA{}) in the direction perpendicular to the monolayers in order to avoid monolayer-monolayer interactions. We fully relaxed both the atoms and the unit cell until forces are well converged below 0.006 eV/\AA{}. Given the importance of stability we also carried out molecular dynamics simulations using the Nose thermostat, and we used a Nose mass of 10.0 Ry f$^{2}$ and a time step of 1 fs \cite{souza2012ab,zanolli2009defective,hobi2010formation}.
We checked the validity of our results by repeating key calculations using the VASP code, which uses the projected augmented wave method (PAW)\cite{blochl1994projector,kresse1999ultrasoft}.

\begin{acknowledgments}
This work has been partially supported by the Project FIS2013-48286-C2-1-P of the Spanish Ministry of Economy and Competitiveness MINECO, the Basque Government under the ETORTEK Program 2014 (nanoGUNE2014), and the University of the Basque Country (Grant No. IT-366-07). TAL acknowledge the grant of the MPC Material Physics Center - San Sebasti\'an. FA-G acknowledge the DIPC for their generous hospitality. We would also like to thank staff at the DIPC computer center (TAL, AA and FA-G).
\end{acknowledgments}

\section*{Supporting Information}

\subsection*{Nose thermostat}

\begin{figure}[thpb]
      \centering
\includegraphics[clip,width=0.5\textwidth,angle=0,clip]{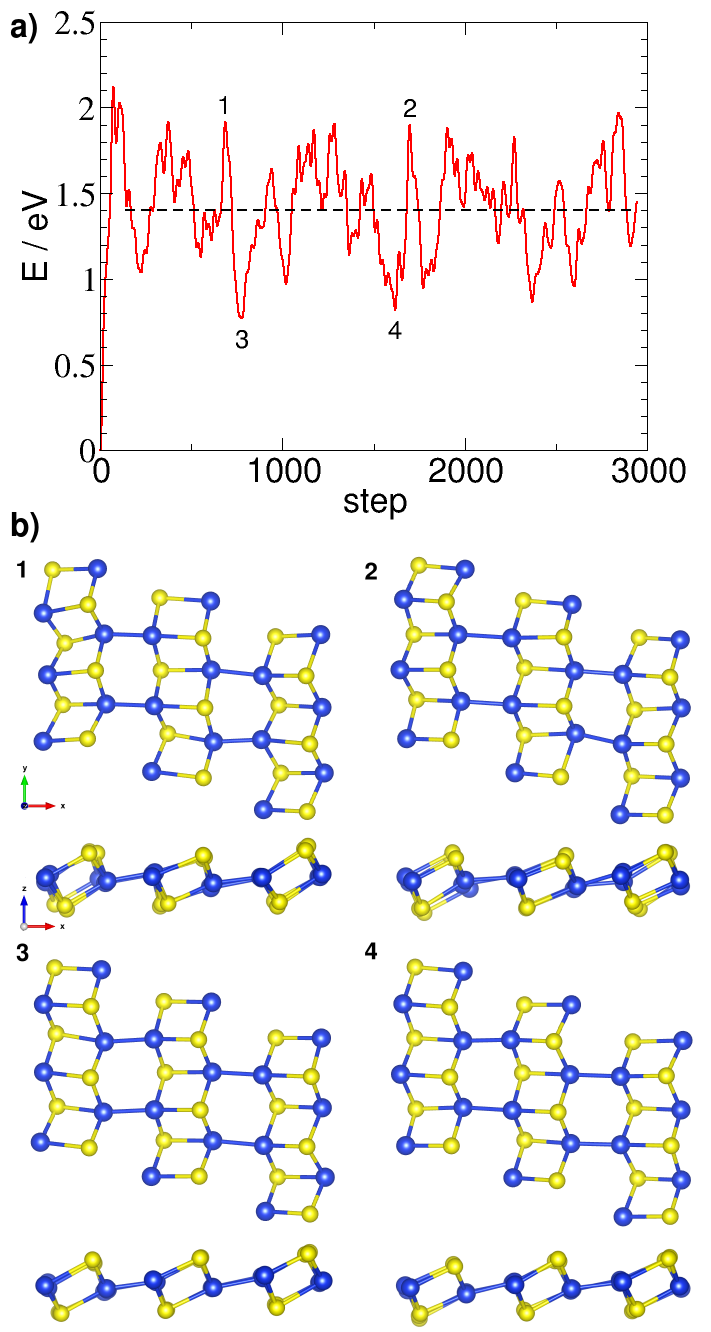}
\caption{\label{figure5}
\textbf{Nose thermostat.}(a) Energy versus time step for the ground state silicon monosulfide monolayer in a Nose thermostat at 300K; (b) snapshots at the four steps marked up in (a).}
\end{figure}

We now expand on the information about the stability of the new ground state monolayer for silicon monosulfide at room temperature. Figure \ref{figure5}(a) shows the energy on each step of the Nose thermostat at 300K. The energy fluctuates slightly accordingly to the small displacements of the atoms with respect to their equilibrium positions, which can be seen in Fig. \ref{figure5}(b). Attending to our simulations we can affirm that the structure is stable at room temperature. We also include two (shown in top and side views) animations that collect the movements of the atoms during the simulations, showing that they just oscillate around their original position.

\subsection*{More SiS geometries}

\begin{figure}[thpb]
      \centering
\includegraphics[clip,width=0.5\textwidth,angle=0,clip]{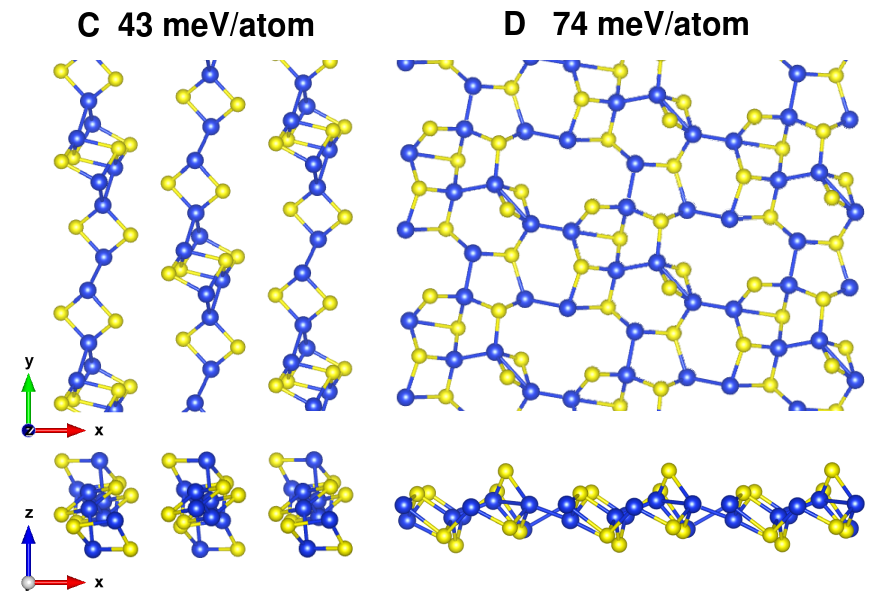}
\caption{\label{figure6}
\textbf{More SiS geometries.} Two other examples of stable monosulfide chain-like structures linked by Si dimers, obtained using large cells. The reference for the energies is referred to the $\gamma$ monolayer.}
\end{figure}

Figure \ref{figure6} shows other silicon monosulfide structures that are also lower in energy than previously reported $\alpha$ and $\beta$ and show plenty of silicon silicon bonds. More structures following this pattern could be identified in future.

%

%

\end{document}